\documentclass[aps,prd]{revtex4}
\bibliography{apsrev}
\begin{document}
\title{Quantum fluctuations of a ``constant'' gauge field}
\author{Antonio Aurilia}
\email{aaurilia@csupomona.edu}
\affiliation{Department of Physics,
 California State Polytechnic University-Pomona}
\author{Euro Spallucci}
\email{spallucci@trieste.infn.it}
\affiliation{Dipartimento di Fisica Teorica,
                                  Universit\`a di Trieste
                          and INFN, Sezione di Trieste}
\date{\today}

\begin{abstract}
It is argued here that the 
quantum computation of the vacuum pressure must take into account
the contribution of zero-point oscillations of a rank-three gauge field.
The field $A_{\mu\nu\rho}$ possesses no radiative degrees of
freedom, its sole function being that of polarizing the vacuum through
the formation of \textit{finite} domains characterized by a non-vanishing, 
constant, but otherwise arbitrary pressure. This extraordinary 
feature, rather unique among quantum fields, is exploited to associate
the $A_{\mu\nu\rho}$ field with the ``bag constant'' of the hadronic 
vacuum, or with the cosmological term in the cosmic case. 
We find that the quantum fluctuations of $A_{\mu\nu\rho}$   
are inversely  proportional to the confinement volume and interpret the result as a 
Casimir effect for the hadronic 
vacuum. With these results in hands and by analogy with the electromagnetic 
and string case, we proceed to calculate the Wilson loop of the 
three-index potential coupled to a ``test'' relativistic bubble. From this 
calculation we extract the static potential between two opposite points on 
the surface of a spherical bag and find it to be proportional to the enclosed 
volume.
   \end{abstract}

\pacs{42.50.Lc, 11.25-w}

\maketitle

\section{Introduction}
It is well known that the cosmological term introduced in General 
Relativity can be expressed as the vacuum expectation value of
the energy-momentum tensor, as one might expect on the basis of 
relativistic covariance
\begin{equation}
\langle\, T^{\mu\nu}\,\rangle = \frac{\Lambda}{8\pi G}\, g^{\mu\nu}\ . \label{tc}
\end{equation}
It is less well known that the same cosmological term can be formulated as
the gauge theory of a rank-three antisymmetric tensor gauge potential
$A_{\mu\nu\rho}$ \cite{uno}, \cite{due}, \cite{tre},\cite{quattro}        with an 
associated field strength 

\begin{equation}
F_{\mu\nu\rho\sigma}=\nabla_{[\, \mu} A_{\nu\rho\sigma\,]}\label{strength}
\end{equation}

invariant under the tensor gauge transformation

\begin{equation}
A_{\mu\nu\rho}\longrightarrow A_{\mu\nu\rho} + \nabla_{[\, \mu} \lambda_{\nu\rho\,]}\ .
\label{abel}
\end{equation}

Indeed, one readily verifies that the classical action 

\begin{equation}
S=-\frac{1}{16\pi G}\,\int d^4x \sqrt{-g}\, R -\frac{1}{2\times 4!}\int d^4x \sqrt{-g}\, 
F^{\lambda\mu\nu\rho}F_{\lambda\mu\nu\rho}\label{scov}
\end{equation}

leads to the familiar Einstein equations in the presence of a cosmological 
term\cite{quattro}, \cite{sei}.
Equation (\ref{tc})  suggests that the cosmological term is associated with the 
zero-point
energy of the cosmic vacuum. Then, in view of the \textit{equivalence} stated above, we are 
naturally led to question the calculability of the zero-point energy due to the quantum 
fluctuations of the $A_{\mu\nu\rho}$-field. At first sight, this may seem as an exercise 
in futility since a \textit{constant} background field, represented by the field strength 
(\ref{strength}),  cannot propagate any physical degree of freedom. However, we shall argue 
in the following sections that there are non-trivial volume effects due to the quantum 
fluctuations of the $A$-field.

Let us switch now from the cosmological case to the hadronic case and 
consider the implications
of quantum vacuum energy in connection with the outstanding problem of color confinement in the
theory of strong interactions. \textit{Somewhat surprisingly, perhaps, the formal connection
between the two extreme cases, cosmological and hadronic, is provided by the same three-index
potential $A_{\mu\nu\rho}$ introduced earlier}. Quantum Chromodynamics is universally 
accepted as
the fundamental gauge theory of quarks and gluons. Equally accepted, however, is the view that
$QCD$ is still poorly understood in the non perturbative regime where the problem of color
confinement sets in. 

On th other hand, the phenomenon of quark confinement is accounted for, as an input, by the
phenomenological ``bag models,'' with or without surface tension \cite{cinque}. In some
such models, it is assumed, for instance,
that the normal vacuum is a color magnetic conductor characterized by an infinite value of the
color magnetic permeability while the interior of the bag, even an empty one, is 
characterized
by a finite color  magnetic permeability. In the interior of the bag the vacuum energy density
acts as a hadronic ``cosmological constant'' originating from zero-point 
energy due to quantum
fluctuations inside the bag. This is a type of Casimir effect for the hadronic vacuum. 
To our knowledge, in spite of the fairly large amount of literature on the subject 
\cite{diciotto}, this effect has never been discussed before in terms of the $A_{\mu\nu\rho}$ 
field.  Ultimately,
the origin of this effect, and therefore of the cosmological bag constant should be 
traced back 
to the fundamental dynamics of the Yang-Mills field. Our suggestion, to be discussed in detail 
in a forthcoming publication, is that the link between
the $A_{\mu\nu\rho}$-field and the fundamental variables of $QCD$ is given by the 
``topological density'' $Tr\mathbf{F}^{\mu\nu} \mathbf{{}^\ast F}_{\mu\nu}$ in $QCD$ through
the specific identification

\begin{equation}
A_{\mu\nu\rho}= {1\over 16\pi^2\, \Lambda^2_{QCD}}\,
Tr\left(\, {\bf A}_{[\,\mu}\,\partial_\nu\, {\bf A}_{\rho\,]} +
{\bf A}_{[\,\mu}\,{\bf A}_\nu \,{\bf A}_{\rho\,]}\,\right)\ . \label{top}
\end{equation}

In support of this identification, notice that a Yang-Mills gauge transformation 
in Eq.(\ref{top}) induces an \textit{abelian} gauge transformation of the type (\ref{abel})

\begin{equation}
\delta A_{\mu\nu\rho}=\frac{1}{g\Lambda_{QCD}}\mathrm{Tr}\left[\, \left(\, 
\mathbf{D}_{[\,\mu}\mathbf{\Lambda}\,\right)\, 
\mathbf{F}_{\nu\rho\,]}\,\right]\equiv 
\frac{1}{g}\partial_{[\,\mu}\,\lambda_{\nu\rho\,]} 
\end{equation}
where $\Lambda_{QCD}$ is the energy scale at which $QCD$ becomes intrinsically 
non-perturbative. \\
Against this background, this paper is the second in a series dealing with the hadronic and
cosmological implications of the vacuum quantum energy associated with the three-index 
potential $A_{\mu\nu\rho}$. In view of the chain of arguments offered above, we shall 
refer to that
field as the ``cosmological field,'' or ``topological field'' depending on the specific 
application under consideration. Some such applications in the cosmological case, in particular
in connection with the problem of dark energy and dark matter in the universe have been 
discussed in our first  paper of the series \cite{sei}. 
Rank three gauge potentials also appear in different sectors of high energy
theoretical physics, e.g., supergravity \cite{due}, cosmology \cite{sette},
both gauge theory of gravity \cite{otto} and of extended objects \cite{nove}. 
As argued above, a central role is played by this kind of gauge field in connection 
with the problem of confinement \cite{sedici}. The present article focuses on the general 
properties of the topological field as an \textit{abelian gauge field of higher rank} but
with an eye on the future discussion of the problem of confinement in $QCD$. The main idea, 
here, is to lay the ground by preparing the tools, both conceptual and technical for that 
discussion. Ultimately, we wish to calculate the Wilson loop for the three-index potential
associated with a bag with a boundary represented by the three-dimensional world history
of a spherical bubble. To our knowledge, this calculation has never been done before and will
pave the way to the future inclusion of fermions in the model. Our calculations are performed
in the euclidean regime and represent a generalization of the more conventional calculations
for the Wilson loop in the case of quantum chromodynamic strings leading to the so called
``area law'' that is taken as a signature of color confinement \cite{dieci} . From the 
Wilson loop
we extract the static potential between two antipodal points on the surface of the bag and
find it to be proportional to the volume enclosed by the surface. This is consistent with
the basic underlying idea of confinement that it would require an infinite amount of energy
to separate the two points. This calculation is performed in Section.4 . As a stepping stone
toward that calculation, we investigate in Section.3 what amounts to the Casimir effect for 
the $A_{\nu\rho\sigma}$ field. Section.2 discusses some of the unique properties of the  
$A_{\mu\nu\rho}$ field that are manifest even at the classical level.   Some concluding remarks
are offered in Sect.5 .

\section{Classically ``trivial'' dynamics}

Rank--three potentials   $A_{\mu\nu\rho}(x)$ were introduced as a
generalization of the elecromagnetic potential and of the Kalb-Ramond potential in string 
theory \cite{undici}, \cite{dodici}, \cite{tredici}.
In the free case, i.e., when there is no interaction with ``matter'', the classical dynamics  
described by the lagrangian density 

\begin{equation}
L_0\equiv {1\over 2\cdot 4!}\,
\left(\, \partial_{[\,\mu}\, A_{\nu\rho\sigma\,]}\,\right)^2\ ,
 \end{equation}
is \textit{exactly solvable}: the field strength $F_{\mu\nu\rho\sigma}\equiv
 \partial_{[\,\mu}\, A_{\nu\rho\sigma\,]}$ that solves the generalized Maxwell
 equations
 
 \begin{equation}
 \partial_\mu\, F^{\mu\nu\rho\sigma}=0
 \end{equation}
 describes a constant background field,  $ F_{\mu\nu\rho\sigma}=f\, 
 \epsilon_{\mu\nu\rho\sigma}$, where $f$ is an arbitrary integration
 constant. In the absence of gravity, such a \textit{classical} constant background field
 has no observable effects and can be rescaled to zero. At the quantum level, we argue 	in the
following, this last statement requires some qualifications. In any case, the 
physical meaning of this background field becomes transparent when
 the $A_{\mu\nu\rho}(x)$ potential is coupled to a rank--three current 
density $J^{\mu\nu\rho} (x)$ with support over the spacetime history of a
relativistic membrane, or $2$-brane, \cite{sette} (~for later convenience, in this paper  we 
work  with euclidean, or Wick rotated, quantities~)

\begin{eqnarray}
L&&= {1\over 2\cdot 4!}\,\left(\, \partial_{[\,\mu}\, A_{\nu\rho\sigma\,]}\,\right)^2
-{\kappa\over 3!}\, J^{\mu\nu\rho} \,  A_{\mu\nu\rho}  \nonumber\\
&&=-{1\over 2\cdot 4!}\, F^{\lambda\mu\nu\rho}\, F_{\lambda\mu\nu\rho}
+{1\over  4!}\, F^{\lambda\mu\nu\rho}\,\partial_{[\,\lambda}\,
 A_{\mu\nu\rho\,]}-{\kappa\over 3!}\, J^{\mu\nu\rho} \,  A_{\mu\nu\rho}
 \label{med}\\
 J^{\mu\nu\rho} \left(\, x\ ; Y\, \right) &&\equiv \int_H 
 \delta\left[\, x-Y\,\right]\, dY^\mu\wedge
dY^\nu\wedge dY^\rho\nonumber\\
&&=\int_{\Sigma } d^3\sigma\, \delta^{4)}\,
\left[\, x-Y\,\right]\,\epsilon^{mnr}\, \partial_m\, Y^\mu\, \partial_n\,
Y^\nu\, \partial_r\, Y^\rho\label{curr}
\end{eqnarray}
where $H $ is the target spacetime image of the world-manifold $\Sigma$
through the embedding $Y: \Sigma\longrightarrow H $.
In the \textit{first order} formulation, $ F_{\lambda\mu\nu\rho}$ and
$A_{\mu\nu\rho}$ are treated as independent variables \cite{quattordici}. However, the
$F$-field equation is algebraic rather than differential, and this provides the 
link between first and second order formulation:

\begin{eqnarray}
&& \frac{\delta L }{\delta F^{\lambda\mu\nu\rho}}=0\longrightarrow
F_{\lambda\mu\nu\rho}=\partial_{[\,\lambda}\, A_{\mu\nu\rho\,]}\\
&& \frac{\delta L }{\delta A^{\mu\nu\rho}}=0\longrightarrow
\partial_\lambda \, F^{\lambda\mu\nu\rho}= \kappa\, J^{\mu\nu\rho} (x)\ .
\label{max}
\end{eqnarray}
The model lagrangian, Eq.(\ref{med}), is the basis for classical and quantum
``membrane dynamics'', CMD and QMD respectively.  Provided that the current
is divergence free, the model is invariant under extended gauge transformations:
 
 \begin{equation}
 \delta A_{\mu\nu\rho}=\partial_{[\,\mu}\,\lambda_{\nu\rho\,]}
 \longleftrightarrow
 \partial_\mu\, J^{\mu\nu\rho} (x)=0\ .  \label{inc}
\end{equation}
The divergence free condition (\ref{inc}) is satisfied whenever the membrane
history has no boundary, which means either:
(i) spatially closed, real membranes, whose world-track is infinitely extended along the 
timelike direction, or
(ii) spatially closed, virtual branes emerging from the vacuum and recollapsing
into the vacuum after a finite interval of proper time \cite{quindici}.\\
To prove that this is the case, let us compute the divergence of the current:

\begin{eqnarray}
\partial_\mu\, J^{\mu\nu\rho} (x)&&=\int_\Sigma d^3\sigma\,\left(\,  
\frac{\partial}{\partial x^\mu}
\delta^{4)}\,\left[\, x-Y\,\right]\,\right)
\,\epsilon^{mnr}\, \partial_m\, Y^\mu\, \partial_n\,
Y^\nu\, \partial_r\, Y^\rho\nonumber\\
&&=\int_\Sigma d^3\sigma\,\left(\,  
\frac{\partial}{\partial Y^\mu}
\delta^{4)}\,\left[\, x-Y\,\right]\,\right)
\,\epsilon^{mnr}\, \partial_m\, Y^\mu\, \partial_n\,
Y^\nu\, \partial_r\, Y^\rho\nonumber\\
&&=\int_\Sigma d^3\sigma\,\left(\,  
\partial_m 
\delta^{4)}\,\left[\, x-Y\,\right]\,\right)
\,\epsilon^{mnr}\,\partial_n\, Y^\nu\, \partial_r\, Y^\rho\nonumber\\
&&=\int_\Sigma d^3\sigma\,\epsilon^{mnr}\partial_m\,
\left(\, \delta^{4)}\,\left[\, x-Y\,\right]
\,\partial_n\, Y^\nu\, \partial_r\, Y^\rho\,\right)\nonumber\\
&&=\int_H  \, d\left(\, \delta^{4)}\,\left[\, x-Y\,\right]\,
 dY^\nu\wedge dY^\rho\,\right)\nonumber\\
&&=\int_{\partial H =\emptyset} \, \delta^{4)}\,\left[\, x-y\,\right]\,
 dy^\nu\wedge dy^\rho=0
 \label{divzero}
\end{eqnarray}

Thus, $\partial_\mu\, J^{\mu\nu\rho} (x)=0\longleftrightarrow  \partial H = \emptyset $. \\ 
If $J$ is divergence free, it can be written as the divergence of
a rank four antisymmetric   \textit{bag current} $K$

\begin{equation}
J^{\mu\nu\rho} (x)\equiv \partial_\lambda\, K^{\lambda\mu\nu\rho}
\label{bulk}
\end{equation}
where 
\begin{equation}
K^{\lambda\mu\nu\rho}(x)\equiv \int_{B} \delta^{4)}\,\left[\,
x-z\,\right]\, dz^\lambda\wedge dz^\mu\wedge dz^\nu \wedge dz^\rho
\end{equation}
and  $H\equiv \partial B$. On the other hand,

\begin{equation}
 dz^\lambda\wedge dz^\mu\wedge dz^\nu \wedge dz^\rho=
 \epsilon^{\lambda\mu\nu\rho}\, d^4z\ ,
\end{equation}
so that one can write $ K^{\lambda\mu\nu\rho}(x)$ as 

\begin{equation}
K^{\lambda\mu\nu\rho}(x)=\epsilon^{\lambda\mu\nu\rho}\,\Theta_{ B}(x)
\label{caratt}
\end{equation}
where
\begin{equation}
\Theta_{ B}(x)=\int_{B} d^4z \,\delta^{4)}\,\left[\,x-z\,\right]
\end{equation}
is the \textit{characteristic function} of the ${ B}$ manifold, i.e., a generalized unit 
step-function:
$\Theta_{ B}\left(\, P  \in {B} \,\right)=1\ ,\quad
\Theta_{ B}\left(\, P  \notin {B} \,\right)=0$.\\
One can also express the bulk-current $K$ in terms of the boundary current $J$
by inverting Eq.(\ref{bulk}): 

\begin{equation}
 \partial_\lambda\, K^{\lambda\mu\nu\rho}=J^{\mu\nu\rho} (x)\longrightarrow
  K^{\lambda\mu\nu\rho}=\partial^{[\,\lambda}\, \frac{1}{\partial^2}\,
  J^{\mu\nu\rho\,]}\ .
\label{bound}
\end{equation}
Now, by solving the Maxwell field equation (\ref{max}), one finds the following
equivalent forms of the classical $F$ field

\begin{eqnarray}
F^{\lambda\mu\nu\rho}=&& f\,\epsilon^{\lambda\mu\nu\rho}
+\kappa\,\partial^{[\,\lambda}\,
\frac{1}{\partial^2}\,J^{\mu\nu\rho\,]}\nonumber\\
=&& \epsilon^{\lambda\mu\nu\rho}\left(\,f  +
\kappa\, \Theta_{ B}(x)\, \right)\label{class}
\end{eqnarray}
where $f$ is, again, the constant solution of the homogeneous equation. 
The presence of the membrane separates spacetime into two regions characterized
by a different value of the energy density and pressure on either side
of the domain-wall \cite{sette}. Thus, the $A$ field produces at most a (constant)
pressure difference between the interior and exterior of a closed
2-brane. However, this special static effect makes the $A$-field a very 
suitable candidate for 
providing a gauge description of the cosmological constant both in 
classical and quantum gravity.\\ 
On the other hand, as we have argued in the Introduction, it is quite
possible that the phenomenon of color
confinement in Quantum Chromodynamics is due to the abelian part of the Yang-Mills field and 
that the long-distance behavior of $QCD$ can be effectively described in terms of 
the rank-three  gauge potential (\ref{top}) associated with the Yang--Mills 
topological density \cite{sedici},\cite{diciasette}.  
Be that as it may, a bag model type of confinement mechanism can be obtained by coupling
$A_{\mu\nu\rho}$ to a membrane current density of the type (\ref{curr}) 
with support on the hadronic bag boundary . That this is the case may be argued even 
at the classical level \cite{uno}. However, our immediate objective here is to link the 
quark bag model 
mechanism of confinement directly to the \textit{quantum} properties of $A_{\mu\nu\rho}$ in
a finite (four) volume.

\section{ Vacuum fluctuations and hadronic Casimir pressure }

In view of our future discussion here and in subsequent articles the message of this section 
needs to be as clear as possible, so we state it at the outset and reiterate it now and then
throughout this section. Suppose there is no closed $2$-brane coupling with $A$.
Then, the free field describes a non-vanishing background energy
associated with the constant field strength $f$ defined in (\ref{class}) with $\kappa=0$. 
Perhaps, this is most simply  understood, physically, in terms of the energy momentum tensor 
derived  from Eq.(\ref{scov}) in the limit of flat spacetime.
 \begin{equation}
 T_{\mu\nu}\equiv -\frac{2}{\sqrt{g}}\frac{\delta S}{\delta g^{\mu\nu}}\vert_{g=\delta}
\longrightarrow \frac{1}{3!}
 F_{\mu\alpha\beta\gamma}\, F_{\nu}^{\alpha\beta\gamma}-
\frac{1}{2\cdot 4!}\,\delta_{\mu\nu}\, F^{\alpha\beta\gamma\delta}\, 
F_{\alpha\beta\gamma\delta}
\end{equation}

From here it follows that 

 \begin{equation} 
T_{\mu\nu}=\frac{f^2}{2}\,\delta_{\mu\nu}\ .
\end{equation}

At first sight, quantizing $A$ seems to be meaningless because there are no dynamical 
degrees of freedom carried by $A$. Against this common misconception we argue
that the \textit{quantum dynamics} of $A$ is non-trivial even in the free
case since a consistent quantization of a ``constant field'' introduces a sort  of 
\textit{volume dynamics.} 
This is best understood in the ``sum over histories approach'' where we have to sum over all 
possible (constant, in our
case) configurations of the field, and weigh each of them with  the usual factor, namely,
  $\exp\left(-\mathrm{euclidean \quad action}\,\right)$. The euclidean action is 
the four volume integral of the lagrangian density evaluated on the given field
configuration. In the case of the $A$-field, the lagrangian density turns out to be constant
over all possible configurations, and the euclidean action is simply:
\textit{ euclidean action $=$ (four volume)$\times$ constant.}
Then, in the limit $V\to\infty$ all quantum fluctuations are frozen and
the value $f=0$ is singled out, as one might reasonably expect in the classical limit.
By reversing the argument, \textit{at the quantum 
level the $A$-field can assume a non vanishing,
constant field strength $F$, only inside a finite volume space(time) region.}
Even if non-dynamical in the usual sense, $A_{\mu\nu\rho}$ plays an active role
anyway: rather than propagating  energy waves, or physical quanta, through spacetime,
 it ``\textit{digs holes into the vacuum}''.
This unique behavior makes $A$ the most appropriate candidate for describing
a ``bubbling vacuum'' in which domains with different vacuum pressure 
endlessly fluctuate in and out of existence.\\
 Mathematically, the above picture of fluctuating virtual 
bubbles can be substantiated in terms of the ``\textit{finite volume}'' partition
functional $Z\left(\, V\,\right)$ 

\begin{eqnarray}
&& Z\left(\, V\,\right)= 
\int \left[\, dF\,\right]\, \left[\, DA\,\right] \,
\exp\left[\, -S_0\left(\,F\ , A\,\right)\,\right]
 \\
&& S_0\left(\,F\ , A\,\right)=\int_V d^4x\,\left[ \frac{1}{2\cdot 4!}
F^2_{\lambda\mu\nu\rho} -\frac{1}{ 4!}F^{\lambda\mu\nu\rho}
\partial_{[\,\lambda}\, A_{\mu\nu\rho\,]}\,\right]\ .
\end{eqnarray}

At this stage,  $V$ represents the characteristic volume of the homogeneous fluctuations of 
the $A$-field. Later we shall discuss the case in which the spacetime region where 
fluctuations take place is bounded by a closed membrane coupled to $A$. 
This whole approach is reminiscent of the Casimir effect for the hadronic vacuum, a 
case-study that has been already widely reported in the literature \cite{diciotto}. 
The novelty of our approach consists in the use of the three-index gauge 
potential, which, to our knowledge, has never been considered before in connection with the
Casimir effect. The main difference lies in the fact that, since the $F$-field is constant
within the region of confinement, it is insensitive to the shape of the boundary, so that the 
resulting Casimir energy density and pressure are also independent of the shape of the 
boundary and are affected only by the size of the volume enclosed.   
In order to substantiate this statement, let us now turn to the technical side of our computation.
\\
Let us start the calculation  of  $Z\left(\, V\,\right)$ from the $A$-integration.
 The $A_{\mu\nu\rho}$ integration measure includes gauge fixing and
Fadeev-Popov ghosts that we will discuss in a short while. 
Before addressing this problem, it is worth observing that the action $S_0$ can also
be written in the form

\begin{equation}
S_0\left(\,F\ , A\,\right)=\int_V d^4x\,\left[\, \frac{1}{2\cdot 4!}
F^2_{\lambda\mu\nu\rho}+
\frac{1}{ 3!}\, A_{\mu\nu\rho} \partial_\lambda\, F^{\lambda\mu\nu\rho}
+\frac{1}{ 3!}\partial_\lambda\,\left(\, A_{\mu\nu\rho} \, 
F^{\lambda\mu\nu\rho} \, \right) 
\right]\ .\label{s02}
\end{equation}
In order to avoid surface terms coming from the total divergence in Eq.(\ref{s02}), we
assume that the volume of quantization has no boundary, for instance it is a four 
sphere.\\
The action $S_0$ is invariant under the gauge transformation

\begin{eqnarray}
&&\delta_\lambda\, A_{\mu\nu\rho}=\partial_{[\,\mu}\,\lambda_{\nu\rho\, ]   }\\
&&\delta_\lambda\, F _ {\mu\nu\rho\sigma  }=0
\end{eqnarray}

and the integration measure over $A$ has to be properly defined in order
to avoid over counting of physically equivalent field configurations. In the
second order formulation, gauge invariance prevents one from inverting the kinetic
operator and from computing the $A$-path integral (~in spite of its gaussian looking form~).
The usual procedure is to break gauge invariance ``by hand'' and compensate
the unphysical degrees of freedom produced by gauge fixing by means of
an appropriate set of ghost fields. In the Lorentz gauge one finds

\begin{equation}
\left[\, DA\,\right]=\left[\, dA\,\right]\,\delta\left[\, \partial_\mu\,
A^{\mu\nu\rho }\,\right]\, \Delta_{FP}
\end{equation}

where the Fadeev-Popov determinant is defined through the gauge variation
of the gauge fixing function

\begin{eqnarray}
\Delta_{FP}&&\equiv \mathrm{det}\left[\, \frac{\delta}{\delta\lambda_{\mu\nu} }
\partial^\rho\,\partial_{[\,\rho}\,\lambda_{\sigma\tau\, ] }\,\right]
\nonumber\\
&&=\mathrm{det}\left[\,\partial^\rho\,\partial_{[\,\rho}\,\delta_\sigma^\mu
\delta_{\tau\,] } ^\nu \,\right]\ .
\end{eqnarray}

The Fadeev-Popov procedure introduces a new gauge invariance which must in turn 
be broken
and compensated until all the unphysical degrees of freedom are removed 
\cite{venti}. This lengthy procedure is necessary in order to perform
perturbative calculations and compute Feynman graphs. However, we are
interested in a non-perturbative evaluation of the path integral. With this
goal in mind, let us remark that in the first order formulation 
$A_{\mu\nu\rho}$ enters
linearly into the action rather than quadratically. In other words, the 
non dynamical nature of $A_{\mu\nu\rho}$ is made manifest in the first order 
formulation, 
where $A_{\mu\nu\rho}$ plays the role of a Lagrange multiplier enforcing the 
classical field equation for $F_{\lambda\mu\nu\rho}$.
Thus, instead of going through all the steps of the Fadeev-Popov procedure,
we split $A_{\mu\nu\rho}$ into  the sum of a Goldstone term $\theta_{\nu\rho}$ 
and a gauge inert part (~modulo a shift by a constant~) 
$\epsilon_{\mu\nu\rho\sigma}\, \partial^\sigma\phi$
\footnote{The dimensions of $\phi$ as defined in Eq.(\ref{31}), are not 
canonical. However,
since $\phi$ is integrated out, we shall leave Eq.(\ref{31}) unchanged.}
\begin{eqnarray}
&& A_{\mu\nu\rho}\equiv  \epsilon_{\mu\nu\rho\sigma}\, 
\partial^\sigma \phi+\partial_{[ \, \mu}\, \theta_{\nu\rho\,] }
\label{31}\\
&& \delta_\lambda \, \phi =0\ ,\qquad \delta_\lambda  \theta_{\nu\rho}=
\lambda_{\nu\rho}\ .
\end{eqnarray}

Accordingly, the functional integration measure becomes

\begin{equation}
\left[\, dA\,\right]= J\,\left[\, d\phi\,\right]\left[\, d\theta
\,\right]
\end{equation}

where $J$ is the functional Jacobi determinant induced by the change of
integration variables (\ref{31}) not to be confused with the
Fadeev-Popov determinant. $J$ reads

\begin{eqnarray}
J &&= \left[\,\mathrm{Det}\left(\,-\partial^2\,\right)\,\right]^{1/2}\times
\left[\,\mathrm{Det}\left(\,-\frac{1}{3!}\epsilon_{\sigma\alpha\beta\gamma}
\partial^\gamma\,\epsilon^{\sigma\alpha\beta\rho}
\partial_\rho\,\right)\,\right]^{1/2}\nonumber\\
&&=\left[\,\mathrm{Det}\left(\,-\partial^2\,\right)\,\right]
\end{eqnarray}

and provides the correct counting of the physical degrees of freedom. Apparently
we introduced two new degrees of freedom: $\theta$ and $\phi$ while from the
classical analysis we expect $A$ to describe a constant background. Let us
show first as $\theta$ drops out form the path integral.\\
The classical action is $\theta$ independent because of gauge invariance 

\begin{equation}
S_0\left(\,F\ , A\,\right)\equiv S_0\left(\,F\ , \phi\,\right)
\end{equation}

and does not provide the necessary damping of gauge equivalent paths.
However, the gauge fixed-compensated integration measure reads

\begin{equation}
\left[\, DA\,\right]\equiv \left[\, d\phi\,\right]\left[\, d\theta
\,\right]\, J\, \delta\left[\, \partial_\mu \partial^{[\, \mu}\,  
\theta^{\nu\rho \,] }\,\right]\,\Delta_{FP}
\label{measure}
\end{equation}

and we can get rid of the gauge orbit volume. Since 

\begin{equation}
\int \left[\, d\theta
\,\right]\delta\left[\, \partial_\mu \partial^{[\, \mu}\,  
\theta^{\nu\rho \,] }\,\right]\,\Delta_{FP}=1
\end{equation}

we obtain a path integral over gauge invariant degrees of freedom
only:

\begin{eqnarray}
&&Z\left(\, V\,\right)= \int \left[\, dF\,\right]\, \left[\, d\phi\,\right]\,
J \, 
\exp\left[ -S_0\left(\,F\ , \phi\,\right)\,\right]
\nonumber\\
&&\ S_0\left(\,F\ , \phi\,\right)\equiv
\int_V d^4x\,\left[\, \frac{1}{2\cdot 4!}
F^2_{\lambda\mu\nu\rho}- \frac{1}{ 3!}\partial_\lambda\,F^{\lambda\mu\nu\rho}
\, \epsilon_{\mu\nu\rho\sigma}\,\partial^\sigma  \,\phi\,   \right]\ .
\end{eqnarray}

Suppose we first integrate over $F$. This is a gaussian integration and we
get

\begin{equation}
Z=\int \left[\, d\phi\,\right]\,
J \, \exp\left[ -\int d^4x\,\frac{1}{2}\phi
\left(-\partial^2\,\right)\left(-\partial^2\,\right)\phi\,
\right]
\end{equation}

Now, if we integrate over the scalar field $\phi$ we see that the contribution
of the $\phi$ field fluctuations exactly cancel the Jacobian because of the
``box squared'' kinetic term:
\begin{equation}
Z= \left[\,\mathrm{Det}\left(\,-\partial^2\,\right)\,\right]\times
\left[\,\mathrm{Det}\left(\,-\partial^2\,\right)^2\,\right]^{-1/2}="1"
\end{equation}

where, the quotation marks is a reminder to the presence of an everywhere 
understood global normalization constant.
Thus, no spurious degrees of freedom have been introduced through through
(\ref{31}).\\
On the other hand, it is interesting to reverse the order of integration and
start with $\phi$ instead of $F$. In this case it is more convenient to
introduce the new integration variable

\begin{equation}
U_{\mu\nu\rho}\equiv \epsilon_{\mu\nu\rho\sigma}\partial^\sigma\phi
\end{equation}

and write the integration measure as

\begin{equation}
\left[\, d\phi\,\right]= \left[\,
dU\,\right]\,\left[\,\mathrm{Det}\left(\,-\partial^2\,\right)\,\right]^{-1/2} 
=J^{-1/2}\, \left[\,dU\,\right]
\end{equation}

Hence, we obtain

\begin{equation}
Z=\int\left[\, dU\,\right] \left[\, dF\,\right]\,  J^{1/2}\,
\exp\left[ -\int_V d^4x\,\left(\,-\frac{1}{2\cdot
4!}F^2_{\lambda\mu\nu\rho}\,
-\frac{1}{ 3!}\partial_\lambda\,F^{\lambda\mu\nu\rho}
\, U_{\mu\nu\rho}\,\right)\,\right]\label{zu}
\end{equation}

We notice that the path-integral is linear in the $U$ variable. 
To integrate over this variable it is convenient to
  rotate, \textit{momentarily}, from euclidean to minkowskian signature
in such a way to reproduce a  path-integral form of the Dirac 
delta-function

\begin{equation}
\int\left[\, dU\,\right] 
\exp\left( -\frac{i}{3!}\int_V d^4x\,\,U_{\mu\nu\rho}\,
\partial_\lambda\,F^{\lambda\mu\nu\rho}\,
\right)=\delta\left[\,
\partial_\lambda\,F^{\lambda\mu\nu\rho}\,\right]
\end{equation}

and then we rotate back to euclidean section. In such a way
the calculation of  $Z\left(\, V\,\right)$ boils down to computing the path 
integral over the field strength configurations that satisfy the ``constraint''
$\partial_\lambda\, F^{\lambda\mu\nu\rho}=0$:

\begin{equation}
Z\left(\, V\,\right)= 
\int \left[\, dF\,\right]\, J^{1/2}\delta\left[\,
\partial_\lambda\,F^{\lambda\mu\nu\rho}\,\right]
\, \exp\left[\,-
\int_V d^4x\, \frac{1}{2\cdot 4!}F^2_{\lambda\mu\nu\rho}\,\right]\ .
\label{vinc} 
\end{equation}
Notice that we are back to the euclidean signature.
Since the constraint is nothing but the classical field equation satisfied by 
$F$, it is easy to implement it since in four dimensions the tensorial 
structure requires that $F_{\lambda\mu\nu\rho}=F(x)\, 
\epsilon_{\lambda\mu\nu\rho}$. Accordingly,
all possible classical solutions are of the form
$F(x)=\mathrm{const.}\equiv f $ where $f$ is an arbitrary parameter. The
path integral is then evaluated by replacing $F$ with its constant value
in the integrand ( and absorbing any field independent quantity in the global
normalization constant ): 

\begin{equation}
Z\left(\,V\ ; f\,\right)
=\exp\left[\,-\frac{1}{2}\, f^2\, V\,\right]
\label{z0}
\end{equation}
which is the standard result available in the literature \cite{tre}. Thus, the
resulting partition function is vanishing in the limit $V\to\infty$
for any value $f\ne 0$. In other words, the only allowed value is $f=0$
giving $Z(V\to\infty)= "1"$. This is the ``trivial
vacuum'' corresponding to a vanishing energy density/pressure.
However, when the volume is finite, one must take into account contributions 
from the \textit{quantum vacuum fluctuations}  of the $F$-field coming from 
all possible , constant, 
values of $f$. Here is where we depart from the conventional formulation of 
the sum over histories approach. Since
$f$ is constant but arbitrary, \textit{the sum over histories amounts to 
integrating over all possible values of $f$ }

\begin{eqnarray}
Z\left(\, V\,\right)&&= \int_{-\infty}^\infty  \frac{df}{\mu_0^2}\,
\int \left[\, dF\,\right]\, J^{1/2}\,
\left[\,\mathrm{Det}\left(\,-\partial^2\,\right)\,\right]^{-1/2}
\delta\left[\,
\,F^{\lambda\mu\nu\rho}- f\,\epsilon^{\lambda\mu\nu\rho}\,\right]
\, \exp\left[\,-
\int_V d^4x\, \frac{1}{2\cdot 4!}F^2_{\lambda\mu\nu\rho}\,\right]
\nonumber\\
&&=\int_{-\infty}^\infty  \frac{df}{\mu_0^2}\,
\exp\left[-\frac{1}{2}\, f^2\,V \, \right]=\sqrt{\frac{2\pi}{V\mu_0^4}}
\label{zv}    
\end{eqnarray}
where $\mu_0$ is a fixed mass scale that is required in order to keep the 
integration measure dimensionless and all the Jacobian factors cancel. The
final result is a ``field independent constant'' which is missing in the
standard formulation. However, this ``constant'' keeps the memory of
$V$ which, in our case, represent the volume where the field fluctuations
takes a non-vanishing value.
Incidentally,  this is the same technique that leads to the correct 
expression for the particle propagator in ordinary quantum mechanics 
\cite{ventuno}.\\
From here we can proceed in two directions. First, we can 
calculate the size of the quantum fluctuations of the $f$-field; second, 
we can derive 
an expression for the vacuum energy density/pressure in the finite volume in 
which the 
quantum fluctuations of the $f$-field are confined.\\
With reference to the first point, since $\Delta f$ is defined as

\begin{equation}
\Delta f\equiv \sqrt{\langle\, f^2\,\rangle -\langle\, f\,\rangle^2    } \label{varf}
\end{equation}
we need to introduce an external source $j$ in order to calculate the average values in 
Eq.(\ref{varf}). By definition

\begin{eqnarray}
&&\langle\, f\,\rangle= -\left(\, \frac{1}{Z(\, f\ , j\,)}\frac{\partial Z(\, f\ , 
j}{\partial j  }\,\right)_{j=0}\label{fmedio}\\
&& \langle\, f^2\,\rangle= \left(\, \frac{1}{Z(\, f\ , j\,)}\frac{\partial^2 Z(\, f\ ,
j}{\partial j^2  }\,\right)_{j=0}\label{f2medio}
\end{eqnarray}
where we use the expression (\ref{zv}) in the presence of an external source

\begin{equation}
Z\left(\, V\,\right)\longrightarrow Z\left(\, V\ ;j\,\right)=
\int_{-\infty}^\infty  \frac{df}{\mu_0^2}\,
\exp\left[-\frac{1}{2}\, f^2\,V -j\,f \, \right]\label{zj}\ .
\end{equation}  

Equations (\ref{fmedio}) and (\ref{f2medio} lead to the following results

\begin{eqnarray}
&&\langle\, f\,\rangle=0\\
&& \langle\, f^2\,\rangle=\frac{1}{V}
\end{eqnarray}

so that the variance of $f$, Eq.(\ref{varf}), is given by

\begin{equation}
\left(\, \Delta f\,\right)^2= \langle\, f^2\,\rangle=\frac{1}{V}\ . 
\label{varfinal}
\end{equation}
The average of the $F$-field turns out to be zero since opposite values of $f$ are weighed 
equally in the partition function (\ref{zv}). However, the final result (\ref{varfinal})
confirms that the quantum fluctuations of the $F$-field are confined in a finite volume, 
with  larger volumes being associated with smaller and smaller fluctuations.\\
Let us now turn back to the promised expression for the vacuum energy density/pressure. 
This follows from the usual definition

\begin{equation}
p\equiv -\frac{\partial}{\partial V}\ln Z(\, V\,)   \ .
\end{equation}
Once we compare it with the explicit expression (\ref{zv}), we find

\begin{equation}
p=\frac{1}{2V}=\frac{1}{2} \langle\, f^2\,\rangle \label{vp}
\end{equation}
which tells us that the Casimir pressure is generated solely by the quantum fluctuations of 
the $F$-field and is inversely proportional to the quantization volume $V$. 
Up until now the volume of confinement has been kept fixed and we have calculated the 
average values of the field $F$ and pressure $p$ inside $V$. At this point we would like to
turn this procedure around and calculate the \textit{average volume} corresponding to
fluctuations with a  preassigned vacuum pressure. Here we face a technical difficulty
since $Z(V)$ behaves as $1/\sqrt V$ and  is therefore non  integrable for large values of 
the argument. In order to get around this difficulty we need to integrate $Z(V)$
over all possible volumes with an appropriate weight factor that plays the role of an
infrared cut-off

\begin{equation}
Z\left(\, \rho_0\,\right)\equiv \int_0^\infty dV\, e^{-\rho_0 V}\, Z\left(\, V\,\right)=
\frac{\pi}{\mu_0^2}\sqrt{\frac{2}{\rho_0}}\ .\label{zrho}
\end{equation}
Using this result we calculate

\begin{equation}
\langle\, V\, \rangle\equiv -\frac{1}{Z\left(\, \rho_0\,\right) }\frac{ \partial Z\left(\, 
\rho_0\,\right)}{\partial\rho_0}=\frac{1}{2\rho_0}\label{vm}
\end{equation}
which illustrates the role of  the infrared cutoff $\rho_0$ and its physical interpretation
as the pressure due to the phenomenological bag constant.

\subsection{Generating Functional}

This subsection has a double purpose: the first is to study the vacuum expectation value of
the energy-momentum tensor as a check on the calculation discussed above; our second purpose 
is to compare the quantum computation of $\langle\,  T_{\mu\nu}\,\rangle $ with its 
classical counterpart already discussed at the beginning of Sect.(3).\\ 
In order to study vacuum expectation values we need to introduce an 
appropriate \textit{external source} coupled to the selected operator and then 
compute the corresponding generating functional.
In our problem the hadronic vacuum pressure and energy density can be extracted from the
expectation value of the energy momentum tensor operator
 
 \begin{equation}
\langle\,  T_{\mu\nu}\,\rangle = 
\langle\,  \frac{1}{3!}
 F_{\mu\alpha\beta\gamma}\, F_{\nu}^{\alpha\beta\gamma}-
\frac{1}{2\cdot 4!}\,\delta_{\mu\nu}\, F^{\alpha\beta\gamma\delta}\, 
F_{\alpha\beta\gamma\delta}\,\,\rangle
\end{equation}

$ T_{\mu\nu}$ being the ``current'' canonically conjugated to the metric 
tensor. Thus, we switch-on a non-trivial background metric $g_{\mu\nu}(x)$

\begin{equation}
S_0\longrightarrow \frac{1}{2\cdot 4!}\int_V d^4x \, \sqrt{g}\,
\,g^{\alpha\beta}\, g^{\mu\gamma}\, g^{\nu\sigma}\, g^{\rho\tau} \, 
F_{\alpha\mu\nu\rho}\, F_{\beta\gamma\sigma\tau}
\end{equation}

where $g\equiv det \, g_{\mu\nu}(x)$. The metric $g_{\mu\nu}(x)$ plays the
role of external source for $ T_{\mu\nu}$, which  means

\begin{equation}
\langle\,  T_{\mu\nu}\,\rangle \equiv \left(\,\frac{2}{\sqrt{g}} 
\frac{\delta}{\delta g^{\mu\nu}(x)}        
   \,\ln\, Z\left[\, g\ ; V\,\right]\, \right)_{g=\delta   }\ .
\end{equation}
Thus, the result is formally the same as in Eq.(\ref{zv}) except for the presence of
$\sqrt{g}$ in the expression of the volume:

\begin{eqnarray}
&& Z\left[\, g\ ; V\,\right]=\left(\, \frac{2\pi}{\mu_0^4\, V[g] }\,\right)^{1/2}\\ 
&&V[g]=\int_V d^4x\sqrt{g}\ ,\quad V[g=\delta]=V\ .
\end{eqnarray}

The same difficulty noted before, namely, the non-integrability of $ Z\left[\, g\ ; 
V\,\right]$ for large $V[g]$ requires that, in order to generate a classical
background pressure in $\langle\,  T_{\mu\nu}\,\rangle $ we consider the regularized
partition function
\begin{equation}
 Z\left[\, g\ ; V\,\right]\longrightarrow Z_{reg}\left[\, g\ ; V\,\right]\equiv
Z\left[\, g\ ; V\,\right]\exp\left(-\rho_0\, V[g]\,\right)\ .\label{zreg}
\end{equation}

Our objective now is to calculate

\begin{equation}
\langle\,  T_{\mu\nu}\,\rangle \equiv \left(\,\frac{2}{\sqrt{g}}\frac{1}{ Z_{reg}\left[\, 
g\ ; V\,\right] } 
\frac{\delta   Z_{reg}\left[\, g\ ; V\,\right]}{\delta g^{\mu\nu}(x)}\, \right)_{g=\delta}\ .
\label{treg}
\end{equation}

Since we have
\begin{equation}
\frac{\delta   Z_{reg}\left[\, g\ ; V\,\right]}{\delta g^{\mu\nu}(x)}=-\frac{1}{2\mu_0^2}
\sqrt{\frac{2\pi}{V[g]}}\sqrt{g}g_{\mu\nu}\left(\, \rho_0 + 
\frac{1}{2V[g]}\,\right)\exp\left(-\rho_0\,V\,\right)\label{deltaz}
\end{equation}

combining equations (\ref{deltaz}), (\ref{zreg}) with the definition (\ref{treg}) we finally
obtain  

\begin{equation}
\langle\,  T_{\mu\nu}\,\rangle \vert_{g=\delta}=\left(\, \rho_0 +
\frac{1}{2V}\,\right)\delta_{\mu\nu}\ . \label{tfinal}
\end{equation}

This final expression of $ T_{\mu\nu}$ confirms the previous calculation of the 
vacuum pressure as consisting of the quantum Casimir pressure superimposed to the 
phenomenological background pressure represented by $\rho_0$. This concludes our discussion of 
the classical and quantum effects due to the three-index potential $A_{\mu\nu\rho}$ in the 
absence of interactions. The coupling to a relativistic test bubble will be the subject   
of next Section.

\section{Hadronic Bags}

In the previous sections we computed the partition function for the  
hadronic vacuum by summing over constant configurations of the $F$-field 
inside finite volume vacuum domains.
The resulting picture is one of a ``bubbling'' ground state in which
\textit{virtual} bags quantum mechanically fluctuate. \\
In this section we wish to study the behaviour of a \textit{real} test bubble immersed in the 
quantum vacuum characterized by the Casimir energy of the $A_{\mu\nu\rho}$ field. 
To begin with,  within the test bubble the $F$-field may 
attain any value as opposed to the exterior (~infinite~) region where its value is zero.\\  
Mathematically,  this new situation corresponds to taking as a new action 
  \begin{equation}
S_0\longrightarrow S_0+\frac{\kappa}{3! }\int d^4x A_{\mu\nu\rho}
J^{\mu\nu\rho}
\end{equation}
where $J^{\mu\nu\rho} $ is given in Eq.(\ref{curr}).\\
The finite volume partition function now reads

\begin{eqnarray}
&& Z\left(\, V\ ; J\,\right)= 
\int \left[\, dF\,\right]\, \left[\, DA\,\right] \,
\exp\left[\, -S\left(\,F\ , A\,\right)\,\right]
 \\
&& S\left(\,F\ , A\,\right)=\int_B d^4x\,\left[ \frac{1}{2\cdot 4!}
F^2_{\lambda\mu\nu\rho} -\frac{1}{ 4!}F^{\lambda\mu\nu\rho}
\partial_{[\,\lambda}\, A_{\mu\nu\rho\,]}-\frac{\kappa}{3!} \, J^{\mu\nu\rho}
A_{\mu\nu\rho}\,  \right]\\
&& V=\int_B d^4x \ .
\end{eqnarray}
 Once again,  let us start the calculation of  $Z\left(\, V\ ; J\,\right)$ from 
the $A$-integration. The only difference with respect to the previous case is that 
a bag is endowed with a non-vanishing boundary. In this case, the total divergenge in 
Eq.(\ref{s02}) may include a surface term defined over $\partial{B}$. 
The most convenient boundary condition is to assume $A$ to be a 
\textit{pure gauge} on $\partial{B}$

\begin{eqnarray}
\frac{1}{ 3!}\int_B d^4x\,
\partial_\lambda\,\left(\, A_{\mu\nu\rho} \, F^{\lambda\mu\nu\rho} \, \right)
 &&=\frac{1}{ 4!}\int_{\partial B} d^3\sigma_{[\, \lambda}\, 
 \partial_{ \mu}\,\lambda_ {\nu\rho\, ] } \,
 \widehat F^{\lambda\mu\nu\rho}\nonumber\\
 &&\equiv \omega\left(\,F\ ,\partial B\,\right) 
\end{eqnarray}

where $\widehat F $ is the  field induced on the boundary by $F$. 
Proceeding in the manner discussed in the previous subsection, we find

\begin{eqnarray}
Z\left(\, V\ ; J\,\right)=&& 
\int \left[\, dF\,\right]\,\delta\left[\,
\partial_\lambda\,F^{\lambda\mu\nu\rho}-\kappa\, J^{\mu\nu\rho}\,\right]
\times\nonumber\\
&& \exp\left[\,-
\int_B d^4x\, \frac{1}{2\cdot 4!}\, F^2_{\lambda\mu\nu\rho}\,\right]\,
\exp\left[\,
-\omega\left(\,F\ , \partial B\,\right)\,\right]\ .
\label{vinc} 
\end{eqnarray}
The surface term does not contribute to the calculation of $ Z\left(\, V\ ; J\,\right)$, after 
integration over  $F$, because of Stoke's theorem, while the effect of the current is to shift 
the constant background value $f$ to $f+\kappa$ within the membrane.  Thus,

\begin{eqnarray}
&& \frac{1}{2\times 4!}\int_B d^4x \, F^{\mu\nu\rho\sigma}\,
 F_{\mu\nu\rho\sigma}=\frac{1}{2\times 4!}\int_B d^4x \,
\left(\,\epsilon^{\mu\nu\rho\sigma} \,f -\kappa\, 
\partial^{[\,\lambda}\,\frac{1}{-\partial^2}\, J^{\mu\nu\rho\,]}
\,\right)^2\nonumber\\
&& =\frac{1}{2}\int_B d^4x \,f^2_{in} - \frac{f\, \kappa}{ 4!}\,
\epsilon_{\mu\nu\rho\sigma} \int d^4x \,
\partial^{[\,\lambda}\,\frac{1}{-\partial^2}\, J^{\mu\nu\rho\,]}
+\frac{\kappa^2}{2\times 3!}\int_B d^4x \,\partial^{[\,\lambda}\,\frac{1}
{-\partial^2}\,
J^{\mu\nu\rho\,]}\,\partial_{[\,\lambda}\,\frac{1}{-\partial^2}\,
J_{ \mu\nu\rho\,]}
\nonumber\\
&&=\frac{1}{2}V \,f_{in}\,\left[\, f_{in} -2\kappa\,\Theta_B(x) \right]+
\frac{\kappa^2}{2\times 3!}\int d^4x \, J^{\mu\nu\rho}\, 
\frac{1}{-\partial^2}\, J_{\mu\nu\rho}\ .
\end{eqnarray}

The final result is obtained after integrating out $f_{in}$:

\begin{equation}
Z\left(\, V \ ; J\,\right)= \sqrt{\frac{2\pi}{\mu^4_0\, V}}\exp\left\{\,
\frac{\kappa^2}{2}V\,\right\} \exp\left(
-\frac{\kappa^2}{2\times 3!}\int_B d^4x \, J^{\mu\nu\rho}\, 
\frac{1}{-\partial^2}\, J_{\mu\nu\rho}\,\right)\ .
\label{wilson}
\end{equation}
The above expression represents the basic generating functional in the interacting case. It 
will be used in the next subsection for the purpose of computing the Wilson loop of the 
$A$-field. Note finally, that for $\kappa=0$ the expression (\ref{wilson}) reduces to the 
free case discussed previously, as it should be. 

\subsection{Wilson factor and the static potential}

In this section we assume that the hadronic manifold $B$ extends indefinitely
along the euclidean time direction and keep the coupling term between $A_{\mu\nu\rho}$ and the 
boundary. Our objective is to determine
 the static potential between pairs of points situated on the boundary
 of the test bubble that we take to be a spherical two-surface of radius
 $R$.  The evolution of the two-sphere in euclidean time is represented by an 
 hyper-cylinder $I\times S^{(2)}$, where $I$ is the interval $0\le t^E \le T$ of euclidean 
time $t^E$. On the two-surface  $   S^{(2)}$  let us ``mark'' a pair of antipodal points and 
follow their (euclidean) time evolution. 
 The two points move along parallel segments of total length $T$.  
 The standard calculation of the static potential between charges moving 
 along  an  elongated rectangular loop, turns, in the
 case under study, into the calculation of the Wilson ``loop'' along the
 hyper-cylinder $I\times S^{(2)}$. The rectangular path is now given by
 the two segments of length $T$ and diameter $2R$ of the sphere at $t^E=0$ and
  $t^E=T$. The corresponding static potential is given by the following
  generalized Wilson integral
 
 \begin{equation}
 V\left(\, R \, \right)\equiv -\lim_{ T\to\infty  }\frac{1}{T}\ln \, W\left[\,
 \partial{B}\right]\ .\label{defv}
 \end{equation}

The path integral calculation of $W\left[\,\partial{B}\right]$ starts
from the finite volume boundary functional, Eq.(\ref{wilson}). The Wilson factor is defined 
as follows

\begin{equation}
 W\left[\, \partial{B}\, \right]\equiv
\langle\, \exp\left[\,
- \frac{\kappa}{3!}\int d^4x \, 
A_{\lambda\mu\nu}\, J^{\lambda\mu\nu}  \,\right]\, \rangle
=\frac{Z\left[\, V\ ; J\right]}{Z\left[\,V\ ; J=0\,\right]}
\end{equation}

where  $V<\infty$ is understood and the limit $V\to \infty$ (along the euclidean time
direction) is performed at the end of the calculations.\\
In order to extract the static potential $V\left(\, R\, \right)$, we compute
the double integral in (\ref{wilson}) for the currents associated to a
pair of antipodal points $P$ and $\overline P$

\begin{eqnarray}
 \int_B d^4x\,
  J^{\mu\nu\rho} \, \frac{1}{\partial^2}\, J_{\,
  \mu\nu\rho}&&=
  \int_{\partial B} \int_{\partial B}
  dy^\mu\wedge dy^\nu\wedge dy^\rho\, \frac{1}{\partial^2}\,
  dy_\mu^\prime \wedge dy_\nu^\prime \wedge dy_\rho^\prime=\nonumber\\
&&=\frac{1}{4\pi^2}\,\int_0^T d\tau\int_T^0 d\tau^\prime
 \int_{ S^{(2)}  } d^2\sigma \int_{ S^{(2)}  } d^2\xi\,\times\nonumber\\
 && y^{\mu\nu\rho}\left(\,\tau\ ,\sigma \,\right)\, \frac{1}{\left[ \, 
  y\left(\, \tau\ ,\sigma\,\right)- y\left( \, \tau^\prime\ ,\xi\,\right) 
  \,\right]^2} \,
  y_{\mu\nu\rho}\left(\, \tau^\prime\ ,\xi \,\right)\,\delta^2\left[\,
  \xi -\sigma\,\right] \nonumber
\end{eqnarray}
where $\left(\, \sigma^1\ ,\sigma^2\,\right)$ and  $\left(\, \xi^1\ ,\xi^2\,\right)$ are two
independent sets of world coordinates on the $S^{(2)}$ manifold. Furthermore, we have 
inserted the explicit form of the scalar Green function and have indicated by

\begin{equation}
y^{\mu\nu\rho}=\epsilon^{abc}\partial_a\, y^\mu\, \partial_a\,
y^\nu\,\partial_b\, y^\rho
\label{tangel}
\end{equation}
the ``tangent elements'' to the world history of the test bubble.
The membrane world manifold is an hypercylinder with euclidean metric
given, in polar coordinates, by
\begin{equation}
ds^2=\gamma_{ab}(\sigma) d\sigma^a d\sigma^b=  d\tau^2 + R^2\left(\, d\theta^2 
+\sin^2\theta\, d\phi^2\,\right)
\end{equation}

where $0\le\phi\le 2\pi$, $0\le\theta\le \pi$, $0\le \tau \le T$.
The embedding in target spacetime is obtained through the equations 

\begin{eqnarray}
&& y^1= R\,\sin\theta\,\sin\phi\\
&& y^2= R\,\sin\theta\,\cos\phi\\
&& y^3= R\,\cos\theta\\
&& y^4=\tau\ .
\end{eqnarray}

Then, with the above choice of coordinates, we find

\begin{eqnarray}
&& y^{\mu\nu\rho}=\partial_{[\, \tau}\, y^\mu\, \partial_\theta\,
y^\nu\,\partial_{\phi\, ] }\, y^\rho \\
&& y^{ijk}\equiv 0\nonumber\\
&& y^{4\, ij}=\partial_{[\, \theta}\, y^i\,\partial_{\phi\, ] }\, y^j\ .
\nonumber
\end{eqnarray}
The explicit expression of the tangent elements $ y^{ijk}$ evaluated at the point $P$ can be 
written as follows

\begin{eqnarray}
&& y^{12}\left(\, \theta\ ,\phi\,\right)\equiv
 \partial_{[\, \theta}\, y^1\,\partial_{\phi\, ] }\, y^2=-R^2\cos\theta\,
\sin\theta\nonumber\\
&& y^{13}\left(\, \theta\ ,\phi\,\right)
\equiv \partial_{[\, \theta}\, y^1\,\partial_{\phi\, ] }\, y^3= 
R^2\sin^2\theta\,\cos\phi\nonumber\\
&& y^{23}\left(\, \theta\ ,\phi\,\right)
\equiv\partial_{[\, \theta}\, y^2\,\partial_{\phi\, ] }\, y^3= 
-R^2\sin^2\theta\,\sin\phi\nonumber\ .
\end{eqnarray}

Then, for the antipodal point $\bar P$ the same expressions become

\begin{eqnarray}
&& y^{12}\left(\,\pi -\theta\ ,\phi +\pi\,\right)=
R^2\cos\theta\,\sin\theta\nonumber\\
&& y^{13}\left(\, \pi -\theta\ ,\phi+\pi\,\right)= 
-R^2\sin^2\theta\,\cos\phi\nonumber\\
&& y^{23}\left(\, \pi -\theta\ ,\phi+\pi\,\right)= 
R^2\sin^2\theta\,\sin\phi\nonumber\ .
\end{eqnarray}

From the above expressions, we explicitly calculate

\begin{eqnarray}
&&\ln W\left[\,{\partial B}\, \right]=\frac{\kappa^2}{48\pi^2}\,
 \int_0^T d\tau  \int_T^0 d\tau^\prime \int_0^\pi d\theta \int_0^{2\pi}d\phi 
 \times\nonumber\\
&&  y^{ij}\left(\,\theta\ ,\phi\,\right)\, \frac{1}
{\left[\, y\left(\,\theta\ ,\phi\,\right) 
-y\left(\,\pi-\theta\ ,\phi+\pi\,\right) \, \right]^2}\,
  y_{ij}\left(\, \pi-\theta\ ,\phi+\pi \,\right)
 \end{eqnarray}

so that

 \begin{eqnarray} 
&&\frac{1}{ 2} y^{ij}\left(\,\theta\ ,\phi\,\right)\, y_{ij}\left(\, \pi-\theta\
,\phi+\pi \,\right)= -R^4\sin^2\theta\\
&&\left[\, y\left(\,\theta\
 ,\phi\,\right) -y\left(\,\pi-\theta\ ,\phi+\pi\,\right)\,\right]^2=
 \left(\, \tau -\tau^\prime\,\right)^2 + 4R^2\ .
 \end{eqnarray}

Therefore the logarithm of $W\left[\,{\partial B}\, \right]$ is

\begin{equation}
\ln W\left[\, \partial B\, \right]=-\frac{\kappa^2\,R^4}{48}\,
 \int_0^T d\tau \int_0^T d\tau^\prime  \, \frac{1}{ \left(\,\tau-
 \tau^\prime \,\right)^2 +4R^2}\ .\label{doppio}
  \end{equation}
We now proceed to calculate the double integral in Eq.(\ref{doppio}):
\begin{eqnarray}
\int_0^T d\tau \int_0^T d\tau^\prime  \, \frac{1}{ \left(\,\tau-
 \tau^\prime \,\right)^2 +4R^2}&&=-\int_0^T d\tau \int_\tau^{\tau -T} du \,
 \frac{1}{  u^2 +4R^2}\ ,\quad u\equiv \tau -\tau^\prime\nonumber\\
&&=-\frac{1}{2R}\int_0^T d\tau\, \int_{\tau/2R}^{(\tau -T)/2R} dy \,
 \frac{1}{ 1 + y^2}\nonumber\\
&&=-\frac{1}{2R}\int_0^T d\tau\,\left[\, \arctan\left(\, \frac{\tau -T}{2R}\, 
\right) -\arctan\left(\, \frac{\tau}{2R}\, \right)\,\right]\nonumber\\
&&\label{doppio1}
 \end{eqnarray}
 
 \begin{eqnarray}
 \int_0^T d\tau\, \arctan\left(\, \frac{\tau -T}{2R}\, \right) &&=
 \int_{-T}^0 ds\, \arctan\left(\, \frac{s}{2R}\, \right)\ ,\quad \tau-T\equiv s
 \nonumber\\
 &&=
 -\int^{T}_0 ds\, \arctan\left(\, \frac{s}{2R}\, \right)\ ,\quad s\rightarrow -s\ .
\label{doppio2} 
\end{eqnarray}
 Putting together equations (\ref{doppio1}) and (\ref{doppio2}) we obtain 

 \begin{equation}
\int_0^T d\tau \int_0^T d\tau^\prime  \, \frac{1}{ \left(\,\tau-
 \tau^\prime \,\right)^2 +4R^2}=\frac{1}{R}\int_0^T d\tau\,\arctan\left(\,
 \frac{\tau}{2R}\, \right)=
 =\frac{T}{R}\arctan\left(\, \frac{T}{2R}\,\right)+ 2R\,\ln\left(\,
  1 +  \frac{T^2}{4R^2}\, \right)
 \end{equation}
 
 which, on account of the definition (\ref{defv}), leads to the final result
 \begin{equation}
 V\left(\, R \, \right)\equiv -\lim_{ T\to\infty  }\frac{1}{T}\ln \, W\left[\,
 \partial{B}\right]=\frac{\pi\, \kappa^2}{96}R^3\ .  \label{V(R)}
 \end{equation}
 
 According to Eq.(\ref{V(R)}) the antipodal points on the spherical membrane of radius $R$ 
 are subject to an attractive potential varying with the volume enclosed by the membrane.

\section{Conclusions}

In this paper we have tried to make a case that the hadronic vacuum represents an ideal
laboratory to test a new approach to the quantum computation of the vacuum pressure in
terms of an antisymmetric, rank-three, tensor gauge field $A_{\mu\nu\rho}$ possibly realized 
in $QCD$ by the collective excitation (\ref{top}) of Yang-Mills fields. A consistent 
formulation of the abelian gauge field $A_{\mu\nu\rho}$  in the sum over histories approach 
requires that the field strength $F$, while
non dynamical in the sense that it propagates no physical quanta, has support over a finite
volume spacetime region, even in the absence of interactions, and gives rise to a Casimir
vacuum pressure that is inversely proportional to the confinement volume. These results
have been confirmed by an explicit computation of the vacuum expectation value of the 
energy-momentum tensor. With such results in hands, we have calculated the Wilson loop of the
three-index potential coupled to a test spherical membrane. From the Wilson factor we have then 
extracted the static potential, Eq.(\ref{V(R)}), between pairs of opposite points on the 
membrane. The ``volume law'' encoded  in Eq.(\ref{V(R)}) is a natural generalization of the 
well known ``area law" for the static 
potential between two test charges (quarks) bound by a chromodynamic string. As a matter of fact, 
 it may be useful to compare the result of equation (\ref{V(R)}) with the more familiar 
result for the Wilson loop of a quark-antiquark pair bounded by a string. In the latter case, the 
integration path is taken to 
be an elongated ( in the euclidean time direction ) rectangle of spatial side $R$. It is
generally assumed that confinement is equivalent to 

\begin{equation}
 W\propto \exp\left(-\sigma\, A\,\right)
\end{equation}
 where $A=TR$ is the area of the rectangle and $\sigma$ is a constant with dimensions of 
(length squared)$^{-2}$. From the definition (\ref{defv}) one extracts a linear potential
between the two test quarks 
\begin{equation}
V(R)=\sigma R \ .\label{linear}
\end{equation} 
The rising of the potential with the distance between charges 
corresponds to the fact that an increasing energy is necessary to separate them. In correspondence
with Eq.(\ref{linear})
we found the expression (\ref{V(R)}) according to which the energy needed to separate 
antipodal points rises as $R^3$.
This cubic law follows from the fact that the two charges under consideration are
located on a  spherical membrane rather than at the endpoints of an open 
string. Note that Eq.(\ref{V(R)}) and Eq.(\ref{linear}) describe the 
same kind of geometric behavior. In both cases the static potential is proportional to the 
``volume'' of the manifold connecting the two test charges. In  Eq.(\ref{linear}), $R$ is
essentially the "linear volume" of the string connecting the pair of test charges. In 
our case, $R^3$ is proportional to the volume of the spherical membrane connecting the two 
antipodal points. Thus, we conclude that in the bag case, confinement is signaled by a ``volume'' 
law extending the string case area  law. It has been noted elsewhere \cite{sedici} that this is the 
exact counterpart, in four spacetime dimensions, of the situation encountered in the two 
dimensional Schwinger model that is widely believed to be the prototype model of quark 
confinement. The precise correspondence of the dynamics of the $A_{\mu\nu\rho}$-field coupled 
to quantum spinor fields and the dynamics of the Schwinger model will be the subject of a 
subsequent article in this series.

\end{document}